\documentstyle[prb,aps,epsf]{revtex}

\begin{document}
\draft

\title{Circuit theory of multiple Andreev reflections in
diffusive SNS junctions: the incoherent case }

\author{E. V. Bezuglyi}
\address{B. Verkin Institute for Low Temperature Physics and
Engineering, Kharkov, 61164 Ukraine \\ NTT Basic Research Laboratories,
Atsugi-shi, Kanagawa, 243-0198 Japan}
\author{E. N. Bratus'}
\address{B. Verkin Institute for Low Temperature Physics and
Engineering, Kharkov, 61164 Ukraine}
\author{V. S. Shumeiko, and G. Wendin}
\address{Chalmers University of Technology, S-41296 Gothenburg,
Sweden}
\author{H. Takayanagi}
\address{NTT Basic Research Laboratories, Atsugi-shi, Kanagawa,
243-0198 Japan}

\date{\today}

\wideabs{
\maketitle

\begin{abstract}

The incoherent regime of Multiple Andreev Reflections (MAR) is studied in
long diffusive SNS junctions at applied voltages larger than the Thouless
energy. Incoherent MAR is treated as a transport problem in energy space by
means of a circuit theory for an equivalent electrical network. The current
through NS interfaces is explained in terms of diffusion flows of electrons
and holes through ``tunnel'' and ``Andreev'' resistors. These resistors in
diffusive junctions play roles analogous to the normal and Andreev reflection
coefficients in OTBK theory for ballistic junctions. The theory is applied to
the subharmonic gap structure (SGS); simple analytical results are obtained
for the distribution function and current spectral density for the limiting
cases of resistive and transparent NS interfaces. In the general case, the
exact solution is found in terms of chain-fractions, and the current is
calculated numerically. SGS shows qualitatively different behavior for even
and odd subharmonic numbers $n = 2\Delta/eV$, and the maximum slopes of the
differential resistance correspond to the gap subharmonics, $eV= 2\Delta/n$.
The influence of inelastic scattering on the subgap anomalies of the
differential resistance is analyzed.
\end{abstract}
}


\narrowtext
\section{Introduction} 
The concept of multiple Andreev reflections (MAR) was first introduced by
Klapwijk, Blonder, and Tinkham \cite{KBT 82} in order to explain the
subharmonic gap structure (SGS) on current-voltage characteristics of
superconducting jun\-c\-tions. The theory was originally formulated for
perfect SNS junctions and then extended to include the effect of resistance
of the SN interface\cite{OTBK} (OTBK theory). Within this approach, the
subgap current transport is described in terms of ballistic propagation of
quasiclassical electrons through the normal metal region, accompanied by
Andreev and normal reflections from specular NS boundaries. During every
passage across the junction, the electrons and the retro-reflected holes gain
energy equal to $eV$, which allows them eventually to escape from the SNS
well. This energy gain results in strong quasiparticle nonequilibrium within
the subgap energy region $|E|<\Delta$.

OTBK theory gives a qualitatively adequate description of dc current
transport in voltage biased SNS junctions; however, its quantitative results
have a rather limited range of applicability. In short ballistic junctions
with length $d$ comparable with or smaller than the coherence length (e.g.,
atomic-size junctions), the quantum coherence of subsequent Andreev
reflections plays a crucial role leading to the ac Josephson effect. It has
been shown that such a coherence also strongly modifies the dc current and
SGS \cite{Arnold,BdGSIS} ({\em coherent} MAR regime). In fact, even in long
ballistic SNS junctions (e.g. 2DEG-based devices), the coherence effects are
important and give rise to resonant structures in the current due to Andreev
quantization. In this respect, the quasiclassical OTBK theory, which does not
include any coherence effects, may be qualified as a model for the {\em
incoherent} MAR regime.

One might expect that impurities could provide the conditions for incoherent
MAR by washing out the Andreev spectrum. However, this is not the case for a
short diffusive junction, where appreciable Josephson coupling gives rise to
coherent MAR.\cite{Zaitsev 98,Zaitsev 90} The electron-hole coherence in the
normal metal holds over a distance of the coherence length $\xi_E =
\sqrt{\hbar{\cal D}/2E}$ from the superconductor (${\cal D}$ is the diffusion
constant). The overlap of coherent proximity regions induced by both SN
interfaces creates an energy gap in the electron spectrum of the normal
metal, which plays the role of the level spacing in the ballistic case. In
short junctions with a wide proximity gap of the order of the energy gap
$\Delta$ in the superconducting electrodes, the phase coherence covers the
entire normal region.

An incoherent MAR regime will occur in long diffusive SNS junctions with a
small proximity gap of the order of Thouless energy $E_{\text {Th}} ={\hbar
\cal D}/d^2\ll \Delta$.\cite{Bezuglyi 99} If the applied voltage is large on
a scale of the Thouless energy, $eV \gg E_{\text {Th}}$, then the coherence
length $\xi_E$ is much smaller than the junction length at all relevant
energies $E \sim \mathop{\mathrm min}(eV,\Delta)$. In this case, the
proximity regions near the SN interfaces become virtually decoupled and the
Josephson oscillations are strongly suppressed. At the same time, as soon as
the inelastic mean free path exceeds the junction length, the subgap
electrons must undergo many incoherent Andreev reflections before they enter
the reservoir. We emphasize that such incoherency is provided by the small
coherence length at large enough voltages, while the intrinsic dephasing
length can be arbitrarily large. In order to describe such an incoherent MAR
regime, one has to operate with the electron and hole diffusion flows across
the junction rather than with ballistic quasiparticle trajectories, and to
consider the Andreev reflections as the relationships between these diffusive
flows.

The first step in extending the OTBK approach to diffusive SNS structures was
taken by Volkov and Klap\-wijk,\cite{Volkov-Klapwijk 92} who derived
recurrence relations between the boundary values of the distribution
functions. In that study, only a weak nonequilibrium was considered, which
implies suppression of MAR by inelastic relaxation. In the present paper, we
focus on the opposite case of strong nonequilibrium in the developed MAR
regime, which results in the appearance of SGS on $I$-$V$ characteristics of
the diffusive SNS junctions.\cite{Experiment} Following the interpretation of
MAR as a transport problem in energy space\cite{Bezuglyi 99,Johansson 98}, we
analyze it by formulating an equivalent network in the spirit of Nazarov's
circuit theory.\cite{CircuitTheory} Within this approach, the
energy-dependent tunnel and Andreev resistances of an equivalent circuit play
roles similar to the normal and Andreev reflection probabilities in OTBK
theory, and the effective voltage source is represented by Fermi reservoirs.

The paper is organized as follows. In Section II, we derive the equations for
incoherent MAR from the general Keldysh equations. In Sections III and IV,
the circuit representation is formulated; some applications are considered in
Section V. The SGS in junctions with resistive interfaces is calculated in
Section VI.  The complete solution of the problem suitable for numerical
calculation of the $I$-$V$ characteristics is obtained in Section VII by
using a chain-fraction technique.\cite{BdGSIS} In Section VIII, we discuss
limitations on the MAR regime imposed by inelastic processes.

\section{Microscopic background} 
\vspace{-3mm}
 The system under consideration consists of a normal channel
($0<x<d$) confined between two voltage biased superconducting electrodes,
with the elastic mean free path $l$ much shorter than any characteristic size
of the problem. In this limit, the microscopic analysis of current transport
can be performed within the framework of the diffusive equations of
nonequilibrium superconductivity\cite{LO} for the $4\times 4$ supermatrix
Keldysh-Green function ${\check{G}}(t_1t_2,x)$:
\begin{equation} \label{Keldysh}
[\check{H},\check{G}]=i \hbar{\cal D}\partial_x \check{J}, \; \check{J}
= \check{G} \partial_x \check{G}, \; {\check{G}}^2=\check{1},
\end{equation}
\begin{equation} \label{Hamiltonian}
\check{H}=\check{1}[i\hbar\sigma_z\partial_t-e\phi(t) +
\hat{\Delta}(t)], \;\hat{\Delta}=\Delta e^{i\sigma_z \chi}i\sigma_y,
\end{equation}
where $\Delta$ is the modulus and $\chi$ is the phase of the order parameter,
and $\phi$ is the electric potential. The Pauli matrices $\sigma_i$ operate
in the Nambu space of $2\times2$ matrices denoted by ``hats'', and the
products of two-time functions are interpreted as their time convolutions.
The junction length $d$ is assumed to be smaller than the inelastic and
phase-breaking lengths, which allows us to exclude the inelastic collisions
from our consideration at this stage; their role will be discussed later. The
electric current $I$ per unit area is expressed through the Keldysh component
$\hat{J}^K$ of the supermatrix current $\check{J}$:
%
\begin{equation} \label{KeldyshCurrent}
I(t) = (\pi\hbar\sigma_N/4e) \mathop{\mathrm{Tr}} \sigma_z \hat{J}^K (tt,x),
\end{equation}
where $\sigma_N$ is the conductivity of the normal metal.

At the SN interface, the supermatrix ${\check{G}}$ satisfies the boundary
condition\cite{Kupriyanov 88}
%
\begin{equation} \label{KeldyshBoundary}
(\sigma_N\check{J})_{\pm0}  = (2R_{SN})^{-1} [{\check{G}}_{-0},{\check{G}}_{+0}],
\end{equation}
where the indices $\pm 0$ denote the right and left sides of the interface
and $R_{SN}$ is the interface resistance per unit area in the normal state,
which relates to, e.g., a Schottky barrier or mismatch between the Fermi
velocities.  Within the model of infinitely narrow potential of the interface
barrier, $U(x)=H\delta(x)$, the interface resistance is related to the
barrier strength $Z=H(\hbar v_F)^{-1}$ as $R_{SN} = 2lZ^2
/3\sigma_N$.\cite{BTK} It has been shown in Ref.\ \onlinecite{Lambert 97}
that Eq.\ (\ref{KeldyshBoundary}) is valid either for a completely
transparent interface ($R_{SN} \rightarrow 0$, ${\check{G}}_{+0}=
{\check{G}}_{-0}$) or for an opaque barrier whose resistance is much greater
than the resistance $R(l)=l/\sigma_N$ of a metal layer with the thickness
formally equal to $l$.

According to the definition of the supermatrix ${\check{G}}$,
%
\begin{equation} \label{DefG}
{\check{G}} = \left( \begin{array}{ccc} \hat{g}^R & \hat{G}^K \\ 0  &
\hat{g}^A \end{array} \right), \; \hat{G}^K = \hat{g}^R \hat{f} -
\hat{f} \hat{g}^A,
\end{equation}
Eqs.\ (\ref{Keldysh}) and (\ref{KeldyshBoundary}) represent a compact form of
separate equations for the retarded and advanced Green's functions
$\hat{g}^{R,A}$ and the distribution function $\hat{f}=f_+ +\sigma_z f_-$.
Their time evolution is imposed by the Josephson relation $\chi(t) = 2eVt$
for the phase of the order parameter in the right electrode (we assume $\chi
= 0$ in the left terminal). This implies that the function
$\check{G}(t_1t_2,x)$ consists of a set of harmonics $\check{G}(E_n,E_m,x)$,
$E_n=E+neV$, which interfere in time and produce the ac Josephson current.
However, when the junction length $d$ is much larger than the coherence
length $\xi_E$ at all relevant energies $E \gtrsim eV$, we may consider
coherent quasiparticle states separately at both sides of the junction,
neglecting their mutual interference and the ac Josephson effect.  Thus, the
Green's function in the vicinity of left SN interface can be approximated by
the solution $\hat{g} =\sigma_z \cosh \theta + i\sigma_y \sinh\theta$ of the
static Usadel equations for a semi-infinite SN structure,\cite{Zaikin} with
the spectral angle $\theta (E,x)$ satisfying the equation 
\begin{equation} \label{SolutionTheta}
\tanh[\theta(E,x)/ 4] = \tanh[\theta_N(E)/ 4] \exp(-x/\xi_E\sqrt{i}),
\end{equation}
with the boundary condition 
\begin{equation} \label{BoundaryTheta}
W\sqrt{i\Delta/E}\sinh (\theta_N-\theta_S) +2\sinh(\theta_N/ 2)=0.
\end{equation}
The indices $S$, $N$ in these equations refer to the superconducting and the
normal side of the interface, respectively.

The dimensionless parameter $W$ in Eq.\ (\ref{BoundaryTheta}), 
\begin{equation} \label{W}
W = {R(\xi_\Delta) \over R_{SN}} = {\xi_\Delta \over d r}, \quad  r =
{R_{SN} \over R_N},
\end{equation}
where $R_N=R(d)=d/\sigma_N$ is the resistance of the normal channel per unit
area, has the meaning of an effective barrier transmissivity for the spectral
functions.\cite{Bezuglyi-Galaiko 99} Note that even at large barrier strength
$Z \gg 1$ ensuring the validity of the boundary conditions Eq.\
(\ref{KeldyshBoundary}),\cite{Lambert 97} the effective transmissivity $W
\sim (\xi_\Delta/l)Z^{-2}$ of the barrier in a ``dirty'' system, $l \ll
\xi_\Delta$, could be large. In this case, the spectral functions are
virtually insensitive to the presence of a barrier and, therefore, the
boundary conditions Eqs.\ (\ref{KeldyshBoundary}) can be applied to an
arbitrary interface if we approximately consider high-transmissive interfaces
with $W \gtrsim \xi_\Delta/l \gg 1$ as completely transparent, $W = \infty$.
For low transmissivity, $W \ll 1$, Eq.\ (\ref{BoundaryTheta}) can be analyzed
within a perturbative approach (see the Appendix). At arbitrary $W$, Eq.\
(\ref{BoundaryTheta}) should be solved numerically.

The distribution functions $f_\pm(E,x)$ are to be considered as global
quantities within the whole normal channel determined by the diffusive
kinetic equations 
\begin{equation}\label{DiffEq}
\partial_x[D_\pm(E,x) \partial_x f_\pm(E,x)] = 0,
\end{equation}
with dimensionless diffusion coefficients 
\begin{mathletters} \label{DiffTheta}
\begin{equation}
D_+ = (1/4)\mathop{\mathrm{Tr}}(1-\hat{g}^R\hat{g}^A) = \cos^2
\mathop{\mathrm{Im}}\theta,
\end{equation}
\begin{equation}
D_- = (1/4)\mathop{\mathrm{Tr}} (1-\sigma_z\hat{g}^R\sigma_z\hat{g}^A)
= \cosh^2 \mathop{\mathrm{Re}}\theta.
\end{equation}
\end{mathletters}

Assuming the normal conductance of electrodes to be much greater than the
junction conductance, we consider them as equilibrium reservoirs with
unperturbed spectral characteristics, $\theta_S = \mathop{\mathrm{Arctanh}}
(\Delta/E)$, and equilibrium quasiparticle distribution, $\hat{f}_S(E) =
f_0(E) \equiv \tanh(E/2T)$. Within this approximation, the boundary
conditions for the distribution functions in Eq.\ (\ref{DiffEq}) at $x = 0$
read 
\begin{equation} \label{BoundaryF}
\sigma_N D_+\partial_x f_+(E,0) = G_+(E)[f_+(E,0)-f_0(E)],
\end{equation}
\begin{equation} \label{BoundaryFz}
\sigma_N D_-\partial_x f_-(E,0)=G_-(E)f_-(E,0),
\end{equation}
where 
\begin{equation} \label{AB}
G_\pm(E) = R_{SN}^{-1}(N_S N_N \mp M_S^\pm M_N^\pm),
\end{equation}
\begin{equation} \label{NM}
N(E)= \mathop{\mathrm{Re}}(\cosh\theta), \;M^+(E)+iM^-(E)=\sinh\theta.
\end{equation}
At large energies, $|E| \gg \Delta$, when the normalized density of states
$N(E)$ approaches unity and the condensate spectral functions $M^\pm(E)$ turn
to zero at both sides of the interface, the conductances $G_\pm(E)$ coincide
with the normal barrier conductance; within the subgap region $|E| < \Delta$,
$G_+(E) = 0$.

Similar considerations are valid for the right NS interface, if we eliminate
the explicit time dependence of the order parameter in Eq.\ (\ref{Keldysh}),
along with the potential of right superconducting electrode, by means of a
gauge transformation\cite{Artemenko 79} 
\begin{equation} \label{Gauge}
{\check{G}}(t_1t_2,x)\! = \!\exp(i\sigma_z eVt_1) \widetilde
{{\check{G}}}(t_1t_2,x) \exp(-i\sigma_z eVt_2).
\end{equation}
As a result, we arrive at the same static equations and boundary conditions,
Eqs.\ (\ref{SolutionTheta})-(\ref{NM}), with $x \rightarrow d-x$, for the
gauge-transformed functions $\widetilde{\hat{g}}(E,x)$ and
$\widetilde{\hat{f}}(E,x)$. Thus, to obtain a complete solution for the
distribution function $f_-$, which determines the dissipative current
%
\begin{equation} \label{Disscurrent}
I = {\sigma_N\over 2e}\int_{-\infty}^\infty dE D_- \partial_x f_-,
\end{equation}
we must solve the boundary problem for ${\hat{f}}(E,x)$ at the left SN
interface, and a similar boundary problem for $\widetilde{\hat{f}}(E,x)$ at
the right interface, and then match the distribution function asymptotics
deep inside the normal region by making use of the relationship following
from Eqs.\ (\ref{DefG}), (\ref{Gauge}): 
\begin{equation}  \label{GaugeF}
{\hat{f}}(E,x) = \widetilde{\hat{f}}(E +\sigma_z eV,x).
\end{equation}

\section{Circuit representation of boundary conditions}
In order for this kinetic scheme to conform to the conventional physical
interpretation of Andreev reflection in terms of electrons and holes, we
introduce the following parameterization of the matrix distribution function,
%
\begin{equation} \label{Defn}
\hat{f}(E,x)= 1-\left(\begin{array}{ccc} n^e(E,x) & 0 \\ 0 & n^h(E,x)
\end{array} \right),
\end{equation}
where $n^e$ and $n^h$ will be considered as the electron and hole population
numbers. Deep inside the normal metal region, they acquire rigorous meaning
of distribution functions of electrons and holes. In equilibrium, the
functions $n^{e,h}$ approach the Fermi distribution. In this representation,
Eqs.\ (\ref{DiffEq}) take the form 
\begin{equation} \label{Diffn}
D_\pm(E,x)\partial_x n_\pm(E,x) = \mbox{const} \equiv
-I_\pm(E)/\sigma_N,
\end{equation}
where $n_{\pm} = n^e\pm n^h$, and they may be interpreted as conservation
equations for the (specifically normalized) net probability current $I_+$ of
electrons and holes, and for the electron-hole imbalance current $I_-$.
Furthermore, the probability currents of electrons and holes, defined as
$I^{e,h} = (1/2)(I_+ \pm I_-)$, separately obey the conservation equations.
The probability currents $I^{e,h}$ are naturally related to the electron and
hole diffusion flows, $I^{e,h} = -\sigma_N \partial_x n^{e,h}$, at large
distances $x \gg \xi_E$ from the SN boundary. Within the proximity region, $x
\lesssim \xi_E$, each current $I^{e,h}$ generally consists of a combination
of both the electron and hole diffusion flows, 
\begin{equation}
I^{e,h}\! =\! -(\sigma_N /2)\!\!\left[ (D_+\!\pm\! D_-)\partial_x n^e\! +\!
(D_+\!\mp\! D_-)\partial_x n^h \right],
\end{equation}
which reflects coherent mixing of normal electron and hole states in this
region.

\begin{figure} [tb]
\epsfxsize=8cm\epsffile{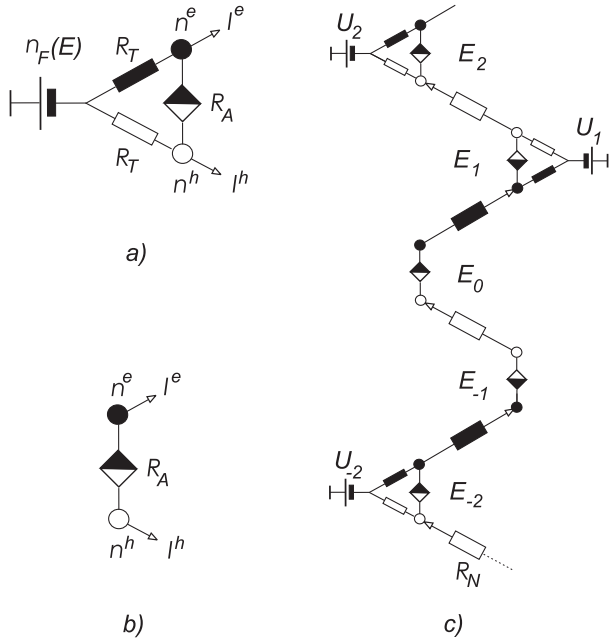} \vspace{0.3cm}
\caption{Elementary equivalent circuits representing bo\-un\-da\-ry
conditions Eq.\ (\ref{BoundaryOhm}) for the electron and hole population
numbers $n^{e,h}(E,0)$ and probability currents $I^{e,h}(E)$, at energies
outside the gap, $|E| > \Delta$ (a), and within the subgap region, $|E| <
\Delta$ (b); equivalent network in ($x,E$)-space for incoherent MAR in SNS
junction (c). Filled and empty symbols stand for electron- and hole-related
elements, respectively; half-filled squares denote Andreev resistors; $U_n =
n_F(E_n)$.} \label{SpaceCircuit}
\end{figure}

In terms of electrons and holes, the boundary conditions in Eqs.\
(\ref{BoundaryF}), (\ref{BoundaryFz}) read 
\begin{equation} \label{BoundaryOhm}
I^{e,h} = G_T(n_F - n^{e,h}) \mp G_A(n^e - n^h),
\end{equation}
where 
\begin{equation} \label{DefCond}
G_T = G_+, \; G_A = (G_- -G_+)/2.
\end{equation}

Each of the equations Eq.\ (\ref{BoundaryOhm}) may be clearly interpreted as
a Kirchhoff rule for the electron or hole probability current flowing through
the effective circuit (tripole) shown in Fig.\ \ref{SpaceCircuit}(a). Within
such an interpretation, the nonequilibrium populations of electrons and holes
$n^{e,h}$ at the interface correspond to ``potentials'' of nodes attached to
the ``voltage source'' --  the Fermi distribution $n_F(E)$ in the
superconducting reservoir -- by ``tunnel resistors'' $R_T(E) = G_T^{-1}(E)$.
The ``Andreev resistor'' $R_A(E) = G_A^{- 1}(E)$ between the nodes provides
electron-hole conversion (Andreev reflection) at the SN
interface.\cite{footnote}

The circuit representation of the diffusive SN interface is analogous to the
scattering description of ballistic SN interfaces: the tunnel and Andreev
resistances in the diffusive case play the same role as the normal and
Andreev reflection coefficients in the ballistic case.\cite{BTK} For
instance, for $|E|>\Delta$ [Fig.\ \ref{SpaceCircuit}(a)], the probability
current $I^e$ is contributed by equilibrium electrons incoming from the
superconductor through the tunnel resistor $R_T$, and also by the current
flowing through the Andreev resistor $R_A$ as the result of hole-electron
conversion. Within the subgap region, $|E|<\Delta$, [Fig.\
\ref{SpaceCircuit}(b)], the quasiparticles cannot penetrate into the
superconductor, $R_T = \infty$, and the voltage source is disconnected, which
results in detailed balance between the electron and hole probability
currents, $I^e = -I^h$ (complete reflection). For the perfect interface,
$R_A$ turns to zero, and the electron and hole population numbers become
equal, $n^e=n^h$ (complete Andreev reflection). The nonzero value of the
Andreev resistance for $R_{SN} \neq 0$ accounts for suppression of Andreev
reflection due to the normal reflection by the interface.

\begin{figure}
\epsfxsize=8cm\epsffile{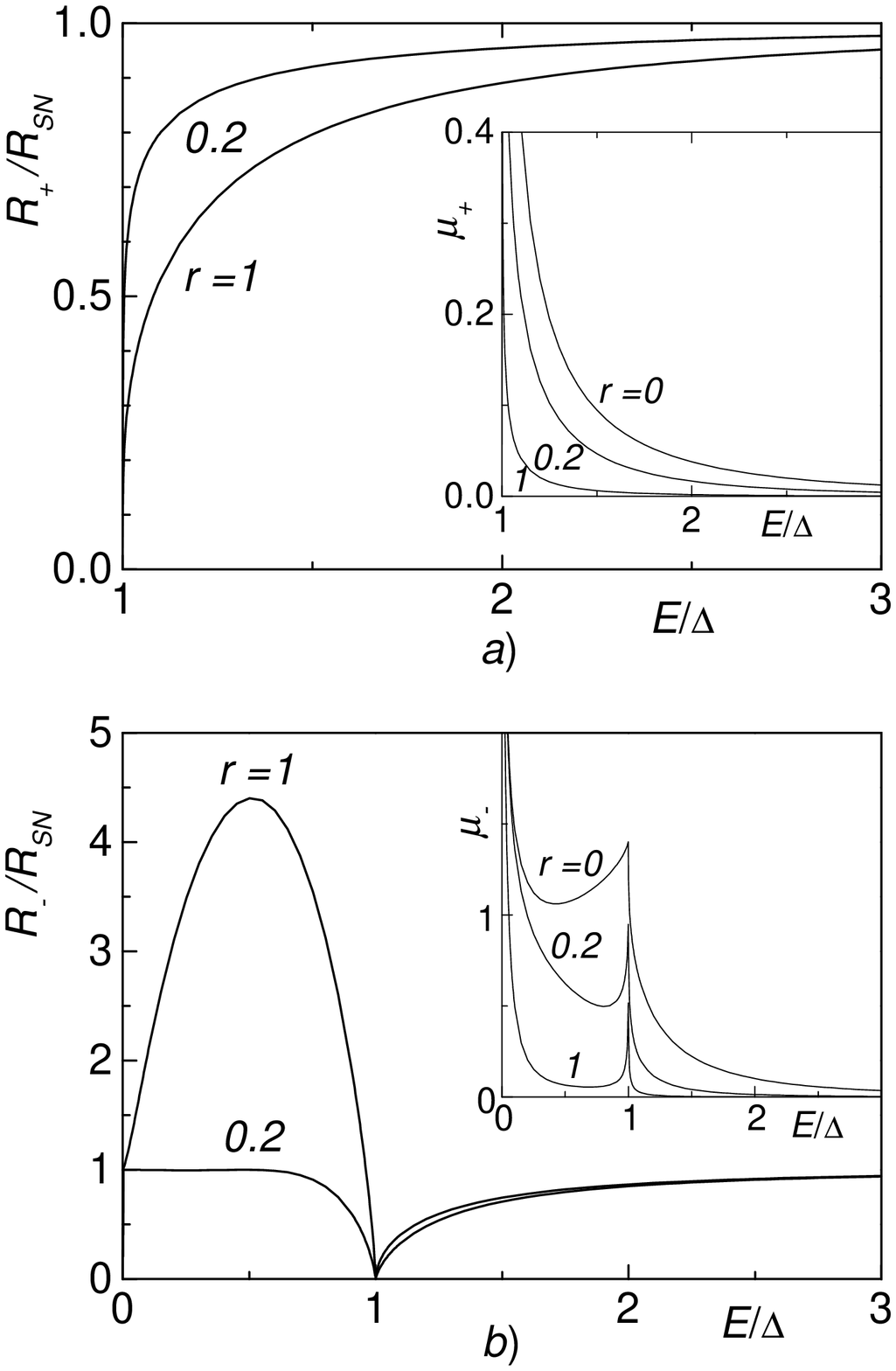}  \vspace{0.3cm} \hspace{-1.5mm}
\epsfxsize=7.7cm\epsffile{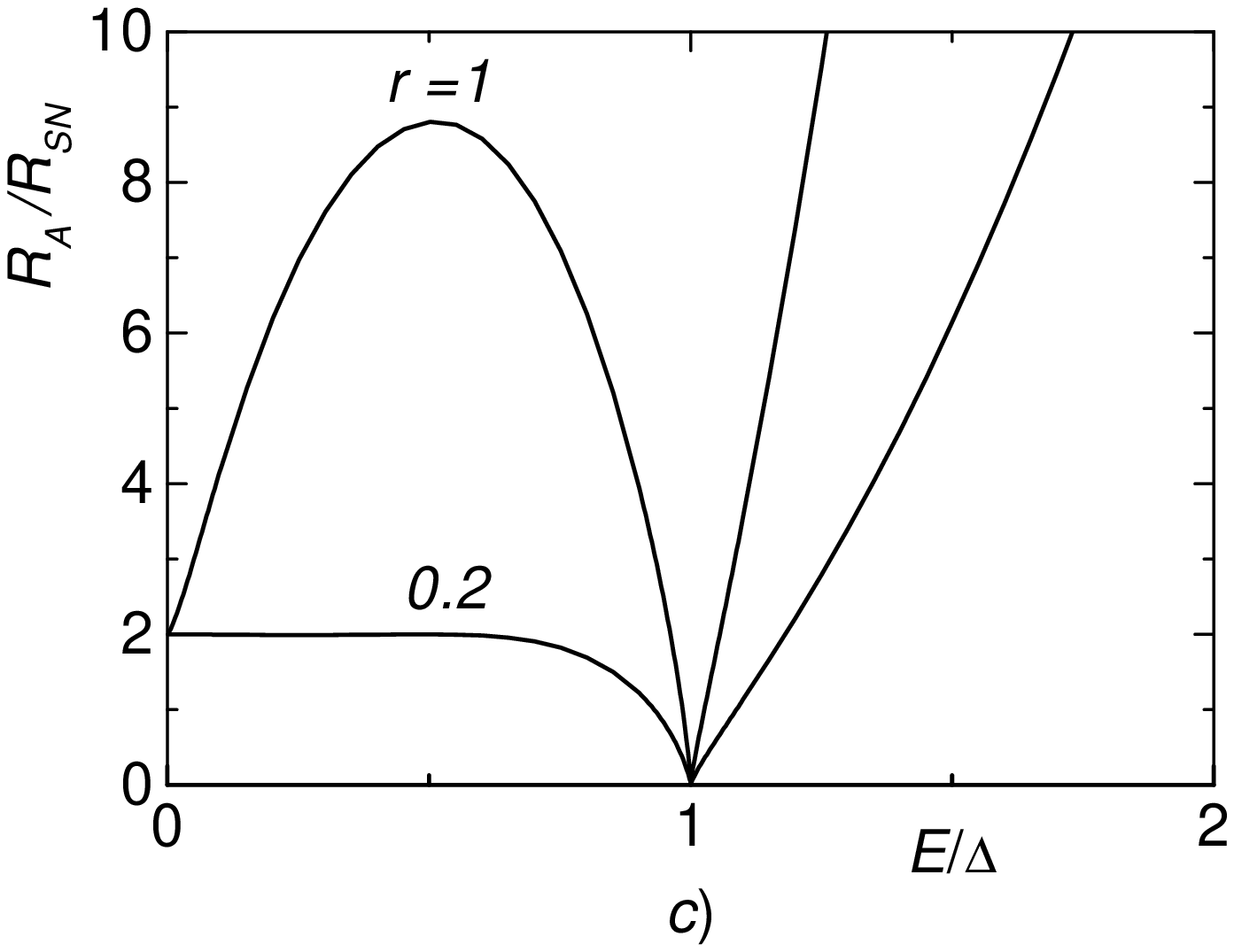} \vspace{0.3cm}
\caption{Energy dependencies of bare interface resistances $R_\pm(E) =
G_\pm^{-1}(E)$ (a,b), bare Andreev resistance $R_A(E)$ (c) and normalized
proximity corrections $\mu_\pm(E)$ [insets in (a) and (b)], for different
values of the resistance ratio $r = R_{SN}/R_N$ and $d/\xi_\Delta = 5$.}
\label{BareResistances}
\end{figure}

Detailed information about the boundary resistances can be obtained from
asymptotic expressions for the bare interface resistances $R_\pm(E)\equiv
G_{\pm}^{-1}(E)$ (see the Appendix) and numerical plots of $R_\pm(E)$ in
Fig.\ \ref{BareResistances}. In particular, $R_\pm(E)$ turns to zero at the
gap edges due to the singularity in the density of states which enhances the
tunneling probability. Furthermore, the imbalance resistance $R_-(E)$
approaches the normal value $R_{SN}$ at $E \rightarrow 0$ due to the
enhancement of the Andreev reflection at small energies, which results from
multiple coherent backscattering of quasiparticles by the impurities within
the proximity region. This property is the reason for the re-entrant behavior
of the conductance of high-resistive SIN systems\cite{Volkov-Klapwijk 92,VZK}
at low voltages. In the MAR regime, one cannot expect any reentrance since
quasiparticles at all subgap energies participate in the charge transport
even at small applied voltage.

The diffusion coefficients $D_\pm$ in Eq.\ (\ref{DiffTheta}) turn to unity
far from the SN boundary, and therefore the population numbers $n^{e,h}$
become linear functions of $x$,
\begin{equation} \label{Asympt}
n^{e,h}(E,x) \approx {\overline n}^{e,h}(E,0) - R_N I^{e,h}(E)x/d.
\end{equation}

\begin{figure}
\epsfxsize=8cm\epsffile{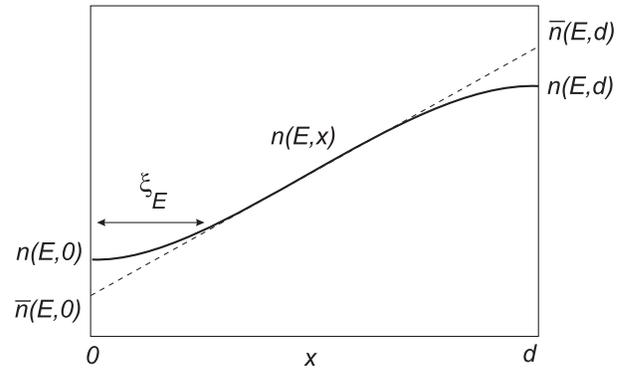} \vspace{0.3cm}
\caption{Qualitative behavior of population numbers within the normal channel
(solid curve). The edge distortions of the linear $x$-dependence of
population numbers, Eq.\ (\ref{Asympt}), occur within the proximity regions.
The difference between the boundary population numbers $n(E,0)$, $n(E,d)$ and
their effective values $\overline n(E,0)$, $\overline n(E,d)$ for true normal
electrons and holes is included in the renormalization of the boundary
resistances, Eq.\ (\ref{RenormAB}).} \label{SketchN}
\end{figure}

This equation defines the renormalized population numbers ${\overline
n}^{e,h}(E,0)$ at the NS interface, which differ from $n^{e,h}(E,0)$ due to
the proximity effect, as shown in Fig.\ \ref{SketchN}. These quantities have
the meaning of the true electron/hole populations which would appear at the
NS interface if the proximity effect had been switched off. It is possible to
formulate the boundary conditions in Eq.\ (\ref{BoundaryOhm}) in terms of
these population numbers by including

\noindent the proximity effect into renormalization of the tunnel and Andreev
resistances. To this end, we will associate the node potentials with
renormalized boundary values $\overline{n}^{e,h}(E,0) =
(1/2)[\overline{n}_+(E,0) \pm \overline{n}_-(E,0)]$ of the population
numbers, where ${\overline n}_\pm(E,0)$ are found from the exact solutions of
Eqs.\ (\ref{Diffn}),
%
\begin{equation} \label{Renormn}
{\overline n}_\pm(E,0) =  n_\pm(E,0) - m_\pm(E)I_\pm(E).
\end{equation}
Here $m_\pm(E)$ are the proximity corrections to the normal metal resistance
at given energy for the probability and imbalance currents, respectively,
%
\begin{mathletters} \label{Defm}
\begin{equation}
m_\pm(E)=\pm R_N (\xi_\Delta/ d)\mu_\pm(E),
\end{equation}
\begin{equation}
\mu_\pm(E) = \int_0^\infty {dx\over \xi_\Delta} \left|
D_\pm^{-1}(E,x)-1\right|>0,
\end{equation}
\end{mathletters}
see the insets in Fig.\ \ref{BareResistances}(a,b). It follows from Eq.\
(\ref{Renormn}) that the same Kirchhoff rules as in Eqs.\
(\ref{BoundaryOhm}), (\ref{DefCond}) hold for ${\overline n}^{e,h}(E,0)$ and
$I^{e,h}(E)$, if the bare resistances $R_\pm$ are substituted by the
renormalized ones, 
\begin{equation} \label{RenormAB}
R_\pm(E) \rightarrow \overline{R}_\pm(E) = R_\pm(E)+ m_\pm(E).
\end{equation}

\begin{figure} [tb]
\epsfxsize=7.5cm\epsffile{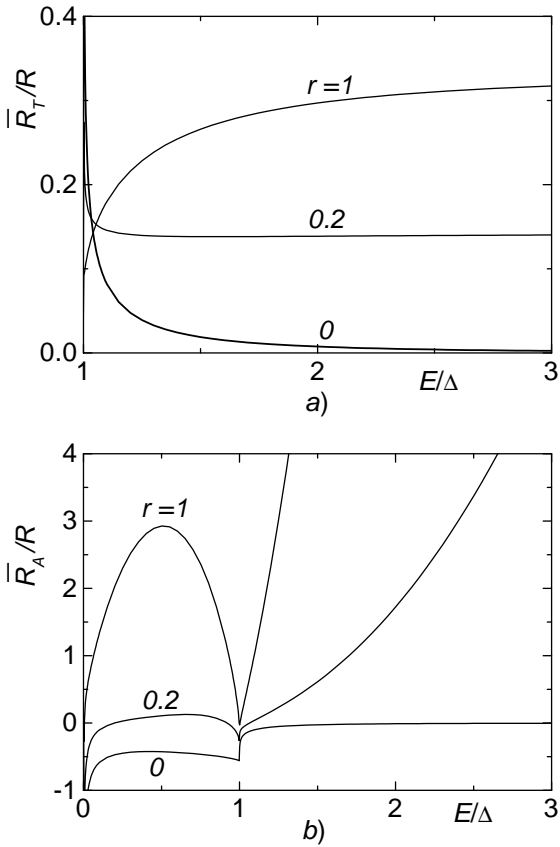} \vspace{0.3cm}
\caption{Energy dependencies of the renormalized interface resistances:
tunnel (probability) resistance $\overline R_T(E)$ (a) and Andreev resistance
$\overline R_A(E)$ (b), for $d=5\xi_\Delta$.} \label{RenormResistances}
\end{figure}

The energy dependence of the renormalized boundary resistances $\overline
R_T(E)$ and $\overline R_A(E)$ is illustrated in Fig.\
\ref{RenormResistances}. In some cases, there is an essential difference
between the bare and renormalized resistances, which leads to qualitatively
different properties of the SN interface for normal electrons and holes
compared to the properties of the bare boundary. Let us first discuss a
perfect SN interface with $R_{SN} \rightarrow 0$. Within the subgap region
$|E| < \Delta$, the bare tunnel resistance $R_T$ is infinite whereas the bare
Andreev resistance $R_A$ turns to zero; this corresponds to complete Andreev
reflection, as already explained. However, the Andreev resistance for normal
electrons and holes, $\overline R_A(E) = 2m_-(E)$, is finite and
negative,\cite{negative} which leads to enhancement of the normal metal
conductivity within the proximity region.\cite{VZK,Stoof} At $|E|>\Delta$,
the bare tunnel resistance $R_T$ is zero, while the renormalized tunnel
resistance $\overline R_T(E) = m_+(E)$ is finite (though rapidly decreasing
at large energies). This leads to suppression of the probability currents of
normal electrons and holes within the proximity region, which is to be
attributed to the appearance of Andreev reflection. Such a suppression is a
global property of the proximity region in the presence of sharp spatial
variation of the order parameter, and it is similar to the over-the-barrier
Andreev reflection in the ballistic systems. In the presence of normal
scattering at the SN interface, the overall picture depends on the interplay
between the bare interface resistances $R_\pm$ and the proximity corrections
$m_\pm$; for example, the renormalized tunnel resistance $\overline R_T(E)$
diverges at $|E|\rightarrow \Delta$, along with the proximity correction
$m_+(E)$, in contrast to the bare tunnel resistance $R_T(E)$. This indicates
complete Andreev reflection at the gap edge independently of the transparency
of the barrier, which is similar to the situation in the ballistic systems
where the probability of Andreev reflection at $|E|=\Delta$ is always equal
to unity.

\section{Assembling MAR networks} 

To complete the definition of an equivalent MAR network, we have to construct
a similar tripole for the right NS interface and to connect boundary values
of population numbers (node potentials) using the matching condition Eq.\
(\ref{GaugeF}) expressed in terms of electrons and holes:
\begin{equation} \label{GaugeN}
n^{e,h}(E,x) = \widetilde{n}^{e,h}(E\pm eV,x).
\end{equation}

Since the gauge-transformed distribution functions $\widetilde{f}_\pm$ obey
the same equations Eq.\ (\ref{DiffEq})-(\ref{NM}), the results of the
previous Section can be applied to the functions $\widetilde{n}^{e,h}(E)$ and
$-\widetilde{I}^{e,h}(E)$ (the minus sign implies that $\widetilde{I}$ is
associated with the current incoming to the right-boundary tripole).  In
particular, the asymptotics of the gauge-transformed population numbers far
from the right interface are given by the equation 
\begin{equation} \label{AsymptGauge}
\widetilde{n}^{e,h}(E,x) \approx \widetilde{\overline{n}}^{e,h}(E,d)
+ R_N \widetilde{I}^{e,h}(E)\left(1-x/d\right).
\end{equation}

After matching the asymptotics in Eqs. (\ref{Asympt}) and (\ref{AsymptGauge})
by means of Eq.\ (\ref{GaugeN}), we find the following relations:
%
\begin{equation} \label{GaugeI}
I^{e,h}(E) = \widetilde{I}^{e,h}(E\pm eV),
\end{equation}
\begin{equation} \label{Matching}
\overline{n}^{e,h}(E,0) - \widetilde{\overline{n}}^{e,h}(E\pm eV,d) = R_N
I^{e,h}(E).
\end{equation}
From the viewpoint of the circuit theory, Eq.\ (\ref{Matching}) may be
interpreted as Ohm's law for the resistors $R_N$ which connect energy-shifted
boundary tripoles, separately for the electrons and holes, as shown in Fig.\
\ref{SpaceCircuit}(c).

\begin{figure} [tb]
\epsfxsize=8.5cm\epsffile{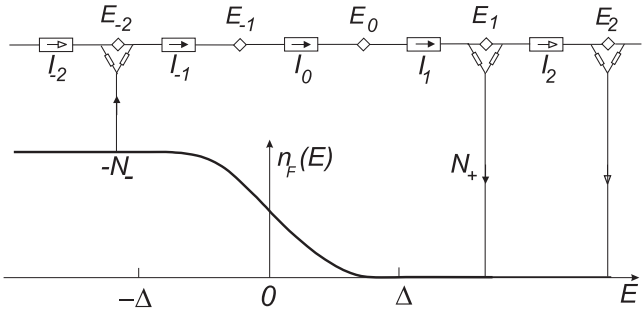} \vspace{0.3cm}
\caption{MAR network of Fig.\ \ref{SpaceCircuit}(c) in energy space. The
nodes outside the gap are connected with the distributed voltage source
$n_F(E)$ (bold curve); the subgap nodes are disconnected from the voltage
source.} \label{EnergyCircuit}
\end{figure}

The final step which essentially simplifies the analysis of the MAR network,
is based on the following observation. The spectral probability currents
$I^{e,h}$ yield opposite contributions to the electric current in Eq.\
(\ref{Disscurrent}), 
\begin{equation} \label{CurrentIeh}
I = {1\over 2e}\int_{-\infty}^{\infty} dE [I^e(E)-I^h(E)],
\end{equation}
due to the opposite charge of electrons and holes. At the same time, these
currents, referred to the energy axis, transfer the charge in the same
direction, viz., from bottom to top of Fig.\ \ref{SpaceCircuit}(c), according
to our choice of positive $eV$. Thus, by introducing the notation $I_n(E)$
for an electric current entering the node $n$ with the energy $E_n = E +
neV$, as shown by arrows in Fig.\ \ref{SpaceCircuit}(c), 
\begin{equation} \label{DefElectric}
I_n(E) = \left\{ \begin{array}{ccc}
I^e(E_{n-1}), & n=2k+1, \\ -I^h(E_n), & n=2k,
\end{array} \right.
\end{equation}
we arrive at an ``electrical engineering'' problem of current distribution in
an equivalent network in energy space plotted in Fig.\ \ref{EnergyCircuit},
where the difference between electrons and holes becomes unessential. The
bold curve in Fig.\ \ref{EnergyCircuit} represents a distributed voltage
source -- the Fermi distribution $n_F(E)$ connected periodically with the
network nodes. Within the gap, $|E_n|<\Delta$, the nodes are disconnected
from the Fermi reservoir and therefore all partial currents associated with
the subgap nodes are equal.

Since all resistances and potentials of this network depend on $E_n=E+neV$,
the partial currents obey the relationship $I_n(E) = I_k[E+(n-k)eV]$ which
allows us to express the physical electric current, Eq.\ (\ref{CurrentIeh}),
through the sum of all partial currents $I_n$ flowing through the normal
resistors $R_N$, integrated over an elementary energy interval $0<E<eV$:
%
\begin{equation} \label{Electric}
I = {1\over 2e}\int_{-\infty}^\infty dE [I_1(E)+I_0(E)] = {1\over
e}\int_0^{eV}dE J(E),
\end{equation}
\begin{equation} \label{SpectralDensity}
J(E) = \sum_{n=-\infty}^{+\infty} I_n(E).
\end{equation}
The spectral density $J(E)$ is periodic in $E$ with the period $eV$ and
symmetric, $J(-E)=J(E)$, which follows from the symmetry of all resistances
with respect to $E$.

As soon as the partial currents are found, the population numbers can be
recovered by virtue of Eqs.\ (\ref{Diffn}), (\ref{BoundaryOhm}),
(\ref{Asympt}), and (\ref{DefElectric}):
\begin{equation} \label{Recovern}
n^{e,h}(E,x) = \overline{n}^{e,h}(E,0) \mp R_N I_{1,0}(E)x/ d,
\end{equation}
\begin{equation} \label{Recovern0}
\overline{n}^{e,h}(E,0) =n_F - {1\over 2} \left[
\overline{R}_+(I_1-I_0) \pm \overline{R}_-(I_1+I_0) \right]
\end{equation}
at $|E|>\Delta$. Within the subgap region, Eq.\ (\ref{Recovern0}) is
inapplicable due to the undeterminacy of product $\overline{R}_+ (I_1-I_0)$.
In this case, one may consider the subgap part of the network as a voltage
divider between the nodes nearest to the gap edges, having the numbers
$-N_-$, $N_+$, respectively, where $N_\pm = \mbox{Int}[(\Delta \mp E)/eV]+1$,
Int($x$) denoting the integer part of $x$. Then the boundary populations at
$|E|<\Delta$ become
\begin{eqnarray} \label{SubgapN}
\overline{n}^{e,h}(E,0) &=& n^{L,R}(E_{\pm N_\pm}) \nonumber \\
&\pm &I_0 \left[ N_\pm
R_N + \sum_{k=1}^{N_\pm-1} R_A(E_{\pm k}) \right],
\end{eqnarray}
where $R,L$ indicate the right (left) node of the tripole, irrespectively of
whether it relates to the left (even $n$) or right (odd $n$) interface. The
physical meaning of $n^{R,L}(E_n)$, however, depends on the parity of $n$:
\begin{equation} \label{Nparity}
n^{R,L}(E_n) = \left\{ \begin{array}{ccc}
\overline{n}^{e,h}(E_n,0), & n=2k, \\
\widetilde{\overline{n}}^{h,e}(E_n,d),  &
n=2k+1. \end{array}\right.
\end{equation}
The values $n^{R,L}$ in Eq.\ (\ref{SubgapN}) can be found from Eq.\
(\ref{Recovern0}) which is generalized for any tripole of the network in
Fig.\ \ref{EnergyCircuit} outside the gap as
\begin{eqnarray} \label{Recover}
n^{R,L}(E_n) =n_F(E_n) - (1/2) \left[
\overline{R}_+(E_n)(I_{n+1}-I_n) \right. \nonumber \\ \left.\pm
\overline{R}_-(E_n)(I_{n+1}+I_n) \right].
\end{eqnarray}

The circuit formalism can be simply generalized to the case of different
transparencies of NS interfaces, as well as to different values of $\Delta$
in the electrodes. In this case, the network resistances become dependent not
only on $E_n$ but also on the parity of $n$. As a result, the periodicity of
the current spectral density doubles: $J(E)=J(E+2eV)$, and, therefore, $J(E)$
is to be integrated in Eq.\ (\ref{Electric}) over the period $2eV$, with an
additional factor $1/2$.

\section{Simple applications} 

\begin{figure}
\epsfxsize=8cm\epsffile{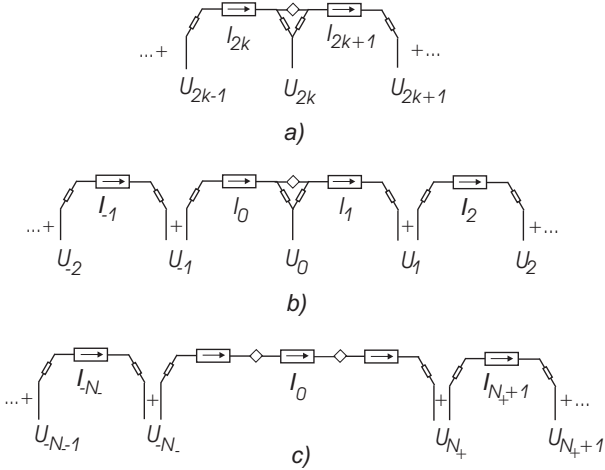}  \vspace{0.3cm}
\caption{Simplified circuits used for calculation of (a) $I$-$V$
characteristic of SNN junction; (b) excess (deficit) current in SNS junction;
(c) $I$-$V$ characteristic of SNS junction with low-transparent interfaces. }
\label{ExampleCircuits}
\end{figure}

A helpful example of an asymmetric junction which allows us immediately to
obtain an analytical solution is given by the SNN structure. The problem of
calculation of its conductivity is inherently static and therefore may be
completely solved for any junction length. If the latter is much larger than
the coherence length, the circuit approach of the previous Section is
applicable without restrictions.  Due to the absence of superconducting
correlations at the right NN interface, odd Andreev resistors are eliminated
and, therefore, the whole network may be split into separate finite circuits
located around even (superconducting) nodes, as shown in Fig.\
\ref{ExampleCircuits}(a); moreover, odd tunnel resistances are to be
considered as normal ones. After some simple algebra, we obtain the sum of
partial currents in each subcircuit,
\begin{equation} \label{PartialSININ}
I_{2k}+I_{2k+1} = {n_F(E_{2k-1})-n_F(E_{2k+1})\over R_N + R_{SN} +
\overline{R}_-(E_{2k})},
\end{equation}
which leads to a well known formula for the $I$-$V$ characteristics of a long
SNN junction:\cite{Volkov-Klapwijk 92}
\begin{eqnarray} \label{SININ}
I = {1\over 2e} \int_0^{2eV} dE \sum_{k=-\infty}^{+\infty}
{n_F(E_{2k-1})-n_F(E_{2k+1})\over R_N + R_{SN} + \overline{R}_-(E_{2k})}
\nonumber \\
= {1\over e} \int_0^\infty dE {n_F(E-eV) -n_F(E+eV)\over R_N +
R_{SN} + \overline{R}_-(E)}.
\end{eqnarray}

If the junction is short enough ($d$ and $\xi_E$ are comparable), one might
naively expect some kind of proximity-induced Andreev scattering at the right
NN interface, followed by MAR and SGS anomalies in the $I$-$V$
characteristic. However, the circuit theory rejects this assumption at once:
since the condensate spectral functions $M^\pm$, Eq.\ (\ref{NM}), disappear
in the normal terminal, the conductivities $G_\pm$ become equal, and the
Andreev channel becomes closed ($G_A = 0$) at the NN interface at any length
of the junction. Thus, the circuit model of charge transport in Fig.\
\ref{ExampleCircuits}(a) remains valid, with a few modifications: (i) the
spectral angle $\theta(E,x)$ has to be found from the Usadel equation for the
finite interval $0<x<d$, (ii) the proximity corrections $m_\pm$ at the left
SN interface are to be expressed through the integrals
\begin{equation}
m_\pm(E) = \pm R_N \int_0^d {dx\over d} \left| D_\pm^{-1}(E,x)-1
\right|,
\end{equation}
instead of Eq.\ (\ref{Defm}), and (iii) odd tunnel resistances $R_{SN}$
should be replaced by the energy-dependent bare resistances $R_+(E)$ [or,
equivalently, $R_-(E)$]. From this point of view, the entire channel
represents a global ``Andreev reflector'' for normal electrons and holes
incoming from the right reservoir.

The next simple application of this circuitry is given by calculation of the
excess (deficit) current in SNS junctions, i.e., the difference $I_{ex} =
I(V) - V/R$ between the currents in superconducting and normal state at large
voltages $V \gg \Delta/e$. Assuming the integration in Eq.\ (\ref{Electric})
to be reduced to the interval $0<E<eV/2$ by making use of the symmetry
$J(E)=J(-E)=J(eV-E)$, we note that the Andreev conductances are negligibly
small for all nodes with $n \neq 0$ ($E_n \gg \Delta$). Thus, the circuit in
Fig.\ \ref{EnergyCircuit} may be split, as in the case of the SNN junction
discussed above, into the chain of separate circuits shown in Fig.\
\ref{ExampleCircuits}(b). The contribution of the central circuit is
described by Eq.\ (\ref{PartialSININ}) with $k=0$, whereas the other parts
are to be considered as normal circuits and represent contribution of
thermally excited quasiparticles:
\begin{equation} \label{PartialExcess}
\sum_{n \neq 0,1}I_n = [1 +n_F(E_1) -n_F(E_{-1})] R^{-1},
\end{equation}
where $R = R_N +2R_{SN}$ is the net normal resistance of the junction. In
summary, we obtain another well-known result,\cite{VZK}
\begin{eqnarray} \label{Excess}
I_{ex} =  {2\over eR} \int_0^{eV/2}\!\!\! dE {n_F(E\!-\!eV)\!
-\!n_F(E+eV)\over R_N + R_{SN} + \overline{R}_-(E)}[R_{SN} \nonumber \\
\!\!\!\!- \overline{R}_-(E)] \approx  {2\over eR} \int_0^\infty dE { R_{SN} -
\overline{R}_-(E) \over R_N + R_{SN} + \overline{R}_-(E)}.
\end{eqnarray}

It is of interest to note that the net resistance $R_T = \overline{R}_+$ for
the probability current never enters final results in these examples and,
therefore, the superconducting modification involves only the imbalance
resistance $\overline{R}_-$. In other words, only the evolution of the
imbalance $n_-$ between the electron and hole populations is relevant for the
charge transport in such cases.

\section{MAR transport}

Proceeding with the analysis of current transport through the SNS junction at
arbitrary voltages, we first discuss the case of low-transparent barriers,
$W\ll 1$. We note that in practice this case is relevant for a wide range of
junctions both with high interface resistance, $R_{SN} \gg R_N$, and
comparatively low interface resistance $R_{SN} \ll R_N$. Indeed, according to
Eq.\ (\ref{W}), the ratio $R_N/R_{SN} = W d/\xi_\Delta$, being proportional
to $W$, contains also the large parameter $d/\xi_\Delta$. Therefore, the
limit $W \ll 1$ covers most of the practically interesting situations, $0 <
R_N/R_{SN} \ll d/\xi_\Delta$, and only the case of very small interface
resistances, $R_N/R_{SN} > d/\xi_\Delta \gg 1$, requires special
consideration.

At $W \ll 1$, the proximity effect is essentially suppressed and the
calculations can be performed on the basis of a simplified model of the
equivalent network, which nevertheless provides a quantitative description of
$I(V)$. Due to the sharp increase in the Andreev resistance at $|E| > \Delta$
[see Fig.\ \ref{RenormResistances}(b)], all Andreev resistors outside the gap
can be excluded, and we arrive at the sequence of subcircuits shown in Fig.\
\ref{ExampleCircuits}(c). The central circuit between the nodes $-N_-$ and
$N_+$ includes $N_++N_-$ normal and $N_++N_--1$ Andreev resistors within the
gap, as well as two tunnel resistors at the circuit edges. The total
resistance $R_\Delta$ of this circuit is
\begin{eqnarray} \label{Rdelta}
R_\Delta(E) = (N_++N_-)R_N+\sum_{-N_-<k<N_+} \overline R_A(E_k)
\nonumber \\ + \overline R_T(E_{N_+}) + \overline R_T(E_{-N_-}),
\end{eqnarray}
and the current $I_0$ through this circuit is given by Ohm's
law:
\begin{equation} \label{SubgapOhm}
I_0(E) = [n_F(E_{-N_-})-n_F(E_{N_+})] / R_\Delta(E).
\end{equation}
Thus, the contribution of this circuit to the current spectral density,
Eq.\ (\ref{SpectralDensity}), is $(N_+ + N_-)I_0$.

The current of thermal excitations is carried by the side circuits ($n
> N_+$, $n \leq -N_-$):
\begin{equation}
I_n = {n_F(E_{n-1}) - n_F(E_n) \over R_N + \overline R_T(E_n) +
\overline R_T(E_{n-1})}.
\end{equation}

From Eq.\ (\ref{SubgapOhm}) we obtain a simple formula at $T \ll
\Delta$:
\begin{equation} \label{ModelCurrent}
I = \int_0^{eV} {dE\over eR_{\text{MAR}}(E)}, \; R_{\text{MAR}}(E) =
{R_\Delta(E) \over N_+ +N_-},
\end{equation}
where $R_{\text{MAR}}(E)$ has the meaning of the effective resistance of the
subgap region for the physical electric current.

\begin{figure}
\epsfxsize=8.5cm\epsffile{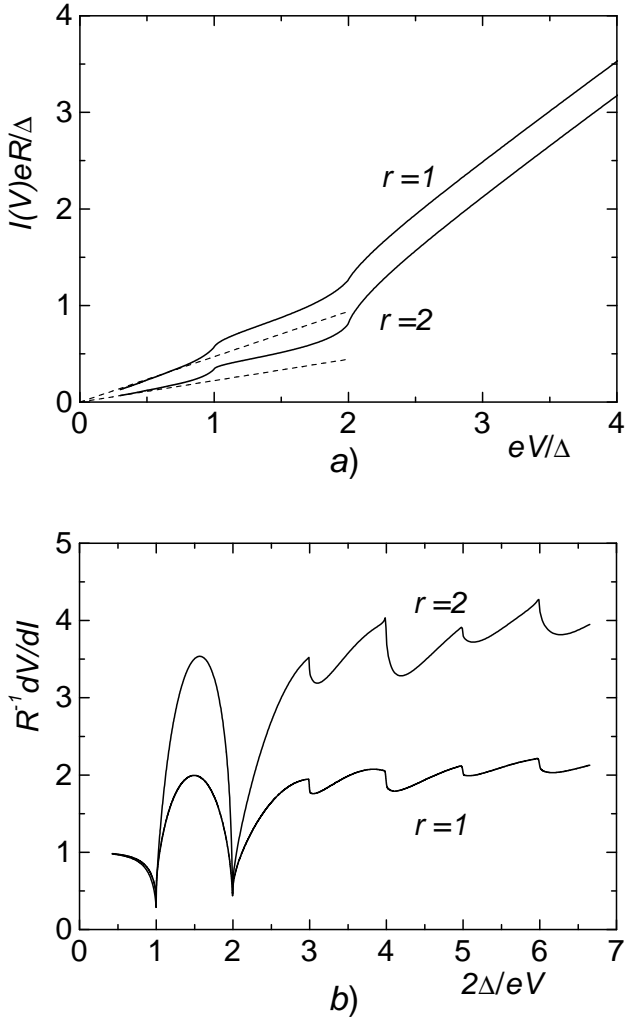}  \vspace{0.3cm}
\caption{$I$-$V$ characteristics (a) and differential resistance vs inverse
voltage (b) of SNS junctions with low-transparent interfaces, $W = 0.1$ and
$W =0.2$, at $d/\xi_\Delta = 5$. Dashed lines denote low-voltage asymptotics
of the $I$-$V$ curves, Eq.\ (\ref{Rlow}). } \label{SimpleIVC}
\end{figure}

In Fig.\ \ref{SimpleIVC} we present the $I$-$V$ characteristics and the
differential resistance vs inverse voltage at zero temperature, calculated
numerically by means of Eq.\ (\ref{ModelCurrent}). The parameter $W$ was
chosen to be equal to $0.1$ and $0.2$ at $d/\xi_\Delta = 5$, which
corresponds to the resistance ratio $r = R_{SN}/R_N$ equal to $2$ and $1$,
respectively.  In our calculation of $\overline R_{T,A}(E)$ in Eq.\
(\ref{Rdelta}), we used the asymptotic Eq.\ (\ref{ApproxR}) for the bare
resistances $R_{T,A}(E)$ at $W \ll 1$, neglecting small proximity corrections
$m_\pm(E) \sim R_N (\xi_\Delta/d) W^2$, Eq.\ (\ref{Perturbm}). The results
are in good agreement with those obtained on the basis of exact calculations
[see further Eq.\ (\ref{T=0})]. The smeared steps in the $I$-$V$
characteristic indicate steplike increase in the number of subgap Andreev
resistors in Eq.\ (\ref{Rdelta}). The sharp peaks (dips) in $dV/dI$ arise
from the rapidly varying contribution of the nodes which simultaneously cross
the gap edges, and therefore both the edge Andreev resistances undergo strong
suppression. A certain contribution also comes from the edge tunnel
resistances which also show singular behavior at $|E| \rightarrow \Delta$.
The peaks are more pronounced for even subharmonics, when the middle Andreev
resistor crosses the Fermi level, $E=0$, and its value is suppressed
simultaneously with the edge Andreev resistors. Careful analysis shows that
at the gap subharmonics, $eV_n = 2\Delta/n$, the second derivative
$d^2V/dI^2$ has sharp maxima.

The magnitude of $dV/dI$ strongly depends on the number of large Andreev
resistors which contribute to $R_{\text{MAR}}$.  At $eV < \Delta$, at least
one Andreev resistor appears far from the ``resonant'' points $E=0,
\pm\Delta$ where $R_A$ sharply decreases. Thus, the net subgap resistance
$R_{\text{MAR}}(E)$ remains large ($\sim R_{SN}/W$) at any energy, which
results in large differential resistance $dV/dI \sim R_{SN}/W$ at these
voltages. In the vicinity of the second subharmonic, $eV \approx \Delta$, the
current transport involves a high-transmissive circuit with three Andreev
resistors located near the resonant points, which yields a much smaller value
of $dV/dI \sim R_{SN}$. The same effect occurs at $eV \approx 2\Delta$ when
the resonant circuit contains two suppressed Andreev resistances at the gap
edges. At $eV > 2\Delta$ the differential resistance is basically determined
by quasiparticles which overcome the energy gap without Andreev reflections,
and it turns to the normal value $R$ at large voltages.

At low voltage, the amplitude of the oscillations of the differential
resistance decreases and the asymptotic value of $dV/dI$ at $V \ll \Delta/e$
can be found analytically from Eqs.\ (\ref{Rdelta}), (\ref{ModelCurrent}), by
replacing the sum in Eq.\ (\ref{Rdelta}) with an energy-independent integral.
As a result we get
\begin{equation} \label{Rlow}
dV/dI(0)\! \approx\!\! R_N \!+\!\! {1\over 2\Delta }\!
\int_{-\Delta}^\Delta\!\!\!\!\! dE R_A(E)\! =\!\! R_N \!+\!\!
{16\sqrt{2}\over 21W} R_{SN}.
\end{equation}

\begin{figure}
\epsfxsize=7.5cm\epsffile{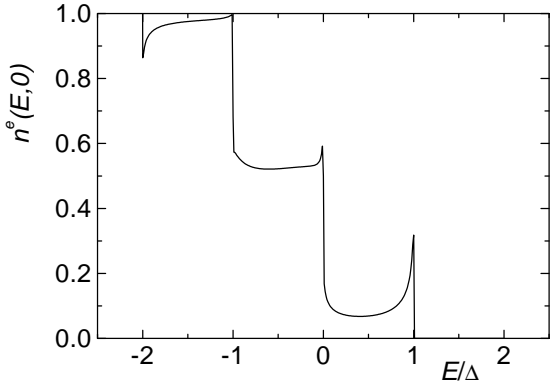}  \vspace{0.3cm}
\caption{Energy dependence of the electron population number $n^{e}(E,0)$ at
the left interface of the SNS junction with $R_{SN} = R_N$ and $d =
5\xi_\Delta$, at voltage $V = \Delta/e$ and $T=0$.} \label{n(E)}
\end{figure}

Since each circuit in Fig.\ \ref{ExampleCircuits}(c) represents a separate
voltage divider, we easily obtain the boundary values of the population
numbers. If the node $n=0$ belongs to the central circuit ($-\Delta-eV < E <
\Delta$ for electrons and $-\Delta < E < \Delta+eV$ for holes), we
have
\begin{eqnarray} \label{nsubgap}
\overline{n}^{e,h}(E,0) = n_F(E_{\pm N_\pm}) \pm {n_F(E_{-N_-}) -
n_F(E_{N_+}) \over R_\Delta(E)} \nonumber \\ \times \left[ R_T(E_{\pm
N_\pm}) + N_\pm R_N + \sum_{k=1}^{N_\pm -1} R_A(E_{\pm k}) \right],
\end{eqnarray}
and, in the opposite case,
\begin{equation} \label{noutgap}
\overline{n}^{e,h}(E,0)\! =\! n_F(E)\!  -\! {R_T(E) [n_F(E)\! -\! n_F(E\pm eV)]
\over R_N + R_T(E) + R_T(E\pm eV)}.
\end{equation}
As follows from Eqs.\ (\ref{nsubgap}), (\ref{noutgap}), at low temperatures,
the energy distribution of quasiparticles within the region $-\Delta-eV < E <
\Delta+eV$ has a steplike form (Fig.\ \ref{n(E)}), which is qualitatively
similar to, but quantatively different from, that found in OTBK
theory.\cite{OTBK} The number of steps increases at low voltage, and the
shape of the distribution function becomes resemblant to a ``hot electron''
distribution with the effective temperature of the order of $\Delta$. This
distribution is modulated due to the discrete nature of the heating mechanism
of MAR, which transfers the energy from an external voltage source to the
quasiparticles by energy quanta $eV$. Since the subgap probability current
$I_0$, Eq.\ (\ref{SubgapOhm}), is strongly suppressed by large subgap
resistance $R_\Delta \sim R_{SN}(N_++N_-)/W \gg R_N$, the spatial variations
of the population numbers, Eqs.\ (\ref{Asympt}), (\ref{Recovern}), are
negligibly small: $n^{e,h}(E,x) - n^{e,h}(E,0) \sim R_N/R_\Delta \ll 1$.

\section{Exact solution} 

In the case of high transmittance of the NS interface, the partial currents
outside the gap noticeably contribute to the net electric current even at $T
= 0$. In this case, the $I$-$V$ curves should be calculated on the basis of
the exact solution of the recurrence relations between partial currents,
%
\begin{equation} \label{Recurrences}
I_{n+1}r_n\! +\! I_{n-1}r_{n-1}\! -\! I_n (\rho_n\! +\! r_{n-1}\! +\! r_n)\!
=\! U_n\! -\! U_{n-1},
\end{equation}
following from the Kirchhoff rules for an infinite network in Fig.\
\ref{EnergyCircuit}. Here $U_n(E) = n_F(E_n)$ and 
\begin{mathletters} \label{RecurrCoeff}
\begin{equation}
\rho_n(E) = R_N + \overline{R}_-(E_{n-1}) + \overline{R}_-(E_n),
\end{equation}
\begin{equation}
r_n(E) = [\overline{R}_+(E_n) -\overline{R}_-(E_n)]/2.
\end{equation}
\end{mathletters}

By analogy with differential equations, we introduce the following ansatz:
%
\begin{equation} \label{Ansatz}
I_n(E)=A_n^+(E)I_n^+(E)+A_n^-(E)I_n^-(E),
\end{equation}
where $I_n^\pm$ are the fundamental solutions of the corresponding uniform
equation, decreasing at $n\rightarrow \pm\infty$, respectively,
\begin{mathletters}  \label{Fundamental}
\begin{equation}
I_n^+(E)=\left\{ \begin{array}{ccc}
S_{n-1,0}, &  n>0,\\ S_{-1,n}^{-1}, & n<0, \end{array} \right.
\end{equation}
\begin{equation}
I_n^-(E)=\left\{\begin{array}{ccc}
P_{n-1,0}^{-1}, & n>0,  \\ P_{-1,n}, & n<0, \end{array} \right.
\end{equation}
\end{mathletters}
$I^+_0=I^-_0=1$. The quantities 
\begin{equation} \label{Products}
S_{mn}(E) = \prod_{j=n}^m s_j(E), \; P_{mn}(E) = \prod_{j=n}^m p_j(E)
\end{equation}
are expressed through the products of chain fractions, $s_n=I^+_{n+1}/
I^+_n$, $p_n = I^-_n / I^-_{n+1}$, defined by the recurrences 
\begin{mathletters}\label{SP}
\begin{equation}
s_n = {r_n \over \rho_{n+1} + r_n + a_{n+1}}, \; p_n = {r_n \over
\rho_n + r_n + b_{n-1}},
\end{equation}
\begin{equation} \label{ab}
a_n=r_n(1-s_n), \; b_n=r_n(1-p_n),
\end{equation}
\end{mathletters}
with the boundary conditions $s_{n\rightarrow +\infty} \rightarrow 0$,
$p_{n\rightarrow -\infty} \rightarrow 0$. Within the gap, $|E_n|<\Delta$,
where $r_n \rightarrow \infty$, the values $s_n$, $p_n$ are equal to $1$, in
accordance with the conservation of the subgap currents mentioned above.

The coefficients $A_n^\pm$ in Eq.\ (\ref{Ansatz}) satisfy an equation
following from Eqs.\ (\ref{Recurrences}), (\ref{Ansatz}),
(\ref{Fundamental}),
\begin{eqnarray} \label{Nonuniform}
[(\rho_{n+1}-r_n s_n^{-1})\delta A^+_{n+1}\ - r_n s_n^{-1} \delta
A^+_n] S_{n,0} + [(\rho_{n+1} \nonumber \\
-r_n p_n)\delta A^-_{n+1}\ - r_n p_n\delta
A^-_n] P_{n,0}^{-1} = U_{n+1} - U_n,
\end{eqnarray}
for $n>0$ (and similar for $n<0$), where $\delta A_n = A_{n+1}-A_n$. The
requirement of cancellation of terms with $\delta A^\pm_{n+1}$ in Eq.\
(\ref{Nonuniform}) allows us to completely determine $A^\pm_n$. This yields
first-order recurrences for $A^\pm_n$ which lead to the formula for partial
currents at $n > 0$,
\begin{equation} \label{Ipositive}
I_n = A_0^+ I_n^+ + A_0^- I_n^- + \sum_{k=1}^n j_k \left(S_{n-1,k} -
P_{n-1,k}^{-1}\right),
\end{equation}
\begin{equation} \label{Fk}
j_n(E) = (U_{n-1}-U_n) /(\rho_n + a_n + b_{n-1}).
\end{equation}

The undeterminacy of $a_n$, $b_{n-1}$ in Eq.\ (\ref{Fk}) within the subgap
region [where $s_n=p_n=1$ and $r_n \rightarrow \infty$, see Eq.\ (\ref{ab})],
is resolved by the recurrences $a_n = \rho_{n+1} + a_{n+1}$, $b_n = \rho_n +
b_{n-1}$ following from Eqs.\ (\ref{SP}) for $|E_n|<\Delta$. These
recurrences are to be continued until the nodes $N_+$ and $-N_-$,
respectively, are reached. As a result, we obtain a convenient representation
for $j_n$ at $(-N_- + 1) \leq n \leq N_+$:
\begin{equation} \label{FkSubgap}
j_n(E) = {U_{n-1}-U_n \over R_\Delta-r_{N_+}(1+s_{N_+})
-r_{-N_-}(1+p_{-N_-})}.
\end{equation}
The effective subgap resistance determined by the denominator in Eq.\
(\ref{FkSubgap}) differs from $R_\Delta$ in Eq.\ (\ref{Rdelta}) by extra
terms describing leakage of the subgap current through the Andreev resistors
outside the gap.

The coefficients $A_0^\pm$ have to provide finite values of $I_n$ in Eq.\
(\ref{Ipositive}) at all $n$; for instance, for $n \rightarrow +\infty$, the
divergent products of $p_j^{-1}$ in $I_n^-$ and $P_{n-1,k}^{-1}$ should
compensate each other: $A_0^- = \sum_{k=0}^{+\infty} j_k P_{k-1,0}$.  A
similar procedure for negative $n$ determines the value of $A_0^+$ and
results in the final formula for the partial current with arbitrary index
$n$,
\begin{eqnarray} \label{FinalIn}
I_n(E) = j_n(E) + \sum_{k=n+1}^{+\infty} j_k(E) P_{k-1,n}(E)\nonumber \\
+\sum_{k=-\infty}^{n-1} j_k(E) S_{n-1,k}(E).
\end{eqnarray}

By making use of the relation $p_{-n}(E)= s_n(-E)$ following from Eq.\
(\ref{SP}) and taking into account the symmetry of all resistances with
respect to $E$, we obtain the net current spectral density,
\begin{eqnarray} \label{J(E)}
J(E) = \sum_{n=-\infty}^{+\infty} \left\{ j_n(E) + \sum_{k=n}^{+\infty}
\left[j_n(E) S_{k,n}(E)
\right. \right.
\nonumber \\
+ j_{-n}(E) S_{k+1,n+1}(-E) ] \Biggr\}.
\end{eqnarray}
At low temperatures $T\ll \Delta$, only the term with $n = 0$ contributes to
the sum in Eq.\ (\ref{J(E)}),
\begin{equation} \label{T=0}
J(E) = j_0(E)\left\{ 1+
\sum_{k=1}^{+\infty} \left[ S_{k-1,0}(E) + S_{k,1}(-E)\right]
\right\}.
\end{equation}

\begin{figure}
\epsfxsize=8.5cm\epsffile{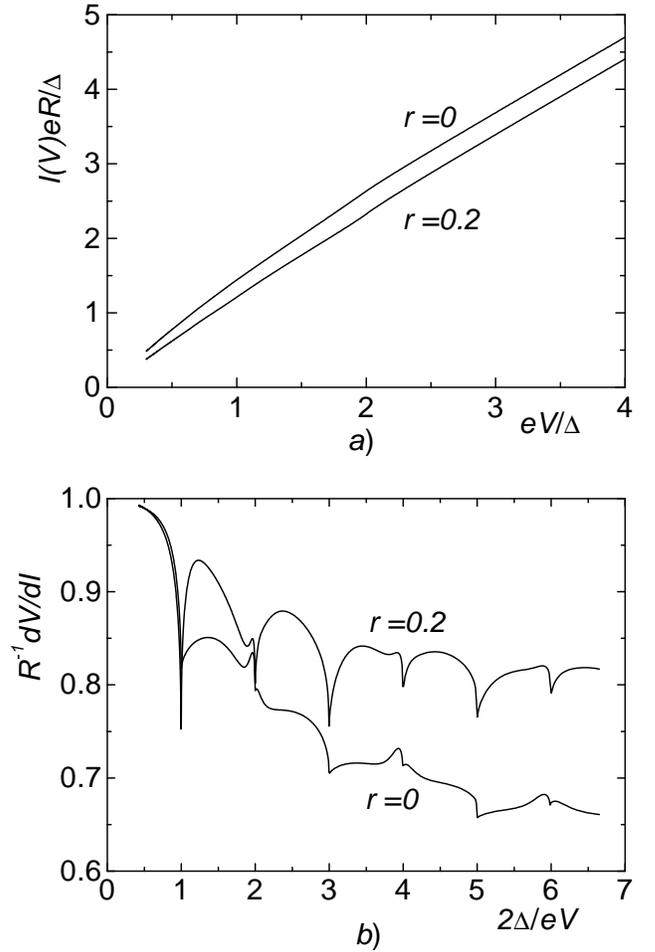}  \vspace{0.3cm}
\caption{$I$-$V$ characteristics (a) and differential resistances vs inverse
voltage (b) of an SNS junction with high-trans\-pa\-rent interfaces.}
\label{ExactIVC}
\end{figure}

Figure \ref{ExactIVC} shows the results of a numerical calculation of $I(V)$
and $dV/dI$ for an SNS junction with high-transmissive interfaces, $W = 1$
and $W \rightarrow \infty$, at zero temperature. In practice, due to rapid
decrease of the coefficients $r_n(E)$ in Eq.\ (\ref{RecurrCoeff}) at large
energies, it is enough to calculate recurrences in Eq.\ (\ref{SP}) starting
from the maximum number $n_{\mathrm{max}}=N_-+2$ and assuming the chain
fractions to be truncated, $s_n = 0$, at $n>n_{\mathrm{max}}$. To avoid
formal singularity in $m_-(E)$ at $E \rightarrow 0$, we introduce a small
dephasing factor $\Gamma$ which provides a cutoff of the coherence length
$\xi_E \rightarrow \sqrt{\hbar{\cal D}/2(E+i\Gamma)}$. The corresponding
dephasing length $l_\phi =\sqrt{\hbar{\cal D}/2\Gamma}$ was chosen equal to
the junction length; the variation of $l_\phi$ is not critical for the fine
structure of $dV/dI$.

Similar to the case of low barrier transmittance, the current transport
through an SNS junction with nearly perfect interfaces can be qualitatively
explained within a simplified model of MAR, where the over-the-barrier
($|E|>\Delta$) Andreev reflection is ignored. Indeed, as follows from Fig.\
\ref{RenormResistances}, at $R_{SN} \ll R_N$, the tunnel resistances
$\overline R_T(E)$ outside the gap are much smaller than the Andreev and/or
normal resistances, except for a narrow region around the gap edges, where
$\overline R_T(E)$ diverges due to complete Andreev reflection. This allows
us to assume all the normal resistors at $|E|>\Delta$ to be connected
directly to the ``voltage source'' $n_F(E)$ and therefore to exclude all
Andreev resistors in Fig.\ \ref{EnergyCircuit} outside the gap. As a result,
we arrive at the sequence of subcircuits shown in Fig.\
\ref{ExampleCircuits}(c), with $\overline R_T = 0$ for side circuits. Thus,
at $T \ll \Delta$, the subgap current may be approximately described by Eqs.\
(\ref{Rdelta}), (\ref{ModelCurrent}), with the tunnel and Andreev resistances
renormalized by the proximity effect.

Within this model, the SGS oscillations in the differential resistance in
Fig.\ \ref{ExactIVC} can be explained in the following way. As the voltage
decreases, the subgap current, which approximately follows Ohm's law,
undergoes an additional suppression in the vicinity of the gap subharmonics,
due to the presence of a high-resistive circuit with two large tunnel
resistors located just at the gap edges. These current steps, being almost
invisible in the $I$-$V$ characteristic, manifest themselves as sharp dips in
$dV/dI$.  At even subharmonics, this effect is partially compensated by the
middle negative Andreev resistor, which rapidly reduces the effective normal
metal resistance due to the increase in the size of the proximity region at
small energies. As a result, the even dips become less pronounced and, as
long as the interface transparency increases, turn into small peaks. At low
voltages, the differential resistance approaches a constant value, which can
be estimated for perfect NS interface by the following expression similar to
Eq.\ (\ref{Rlow}):
\begin{eqnarray} \label{Rperfect}
dV/dI(0) \approx R_N \left[ 1- 2{\xi_\Delta\over d} \int_{-\Delta}^\Delta
{dE\over 2\Delta} \mu_-(E)\right] \nonumber \\ = R_N \left( 1-
2.64\xi_\Delta/ d\right), \; R_{SN} = 0.
\end{eqnarray}

Unlike the ballistic SNS junction,\cite{KBT 82,Arnold} but similar to short
diffusive constriction,\cite{Zaitsev 98} the SGS survives at zero temperature
even for perfect NS interfaces. In this case, the SGS occurs due to coherent
impurity scattering of quasiparticles inside the proximity region, with the
amplitude approximately proportional to the characteristic length
$\xi_\Delta$ of this region. If we neglect the proximity corrections ($m_\pm
\rightarrow 0$), the SGS disappears, along with the excess current, and the
$I$-$V$ characteristic shows perfect Ohmic behavior.\cite{Bezuglyi 99} Thus,
we conclude that in the both cases of resistive and transparent NS
interfaces, SGS appears due to the normal electron backscattering in the
proximity region. This formally corresponds to the finite value of the
renormalized Andreev resistance of the interface.

\section{Role of inelastic scattering}

Since the relative SGS amplitude increases along with the NS barrier strength
(though the current itself decreases), one might expect systems with
high-resistive interfaces to be more favorable for the experimental
observation of SGS. However, there exists an intrinsic restriction for the
effect: to provide strong nonequilibrium of the subgap quasiparticles
inherent to MAR, the time $\tau_d$ of their diffusion through the whole MAR
staircase, from $-\Delta$ to $\Delta$, must be smaller than the inelastic
relaxation time $\tau_\varepsilon$. The value of $\tau_d$ can be estimated as
the time for diffusion over an effective length $(2\Delta/eV)d$. At low
interface resistance, $R_{SN} \ll R_N$, the diffusion rate is basically
determined by the impurity scattering:  $\tau_d(eV) \sim (2\Delta/eV)^2
d^2/{\cal D}$. For high interface resistance, $R_{SN} \gg R_N$, the large
Andreev resistance $R_A \sim R_{SN}/W$ present bottleneck which renormalizes
the diffusion coefficient in $\tau_d$ by a small factor $WR_N/R_{SN}$. In
this way, the level of nonequilibrium of the subgap quasiparticles is
controlled by the parameter
\begin{equation} \label{WE}
W_\varepsilon = {\tau_\varepsilon \over \tau_d(2\Delta)}= {R_N \over
R_N + R_{SN}/W} E_{\mathrm{Th}} \tau_\varepsilon,
\end{equation}
which must be large enough to allow observation of at least a few
subharmonics in $I(V)$, i.e., the condition $W_\varepsilon > 1$ determines
the lower boundary for the barrier transparency. An analogous estimate for
the inelastic parameter, with the barrier resistance $R_T \gg R_N$
substituting for the Andreev resistance $R_{SN}/W$ in Eq.\ (\ref{WE}), was
obtained in Ref.\ \onlinecite{Bezuglyi 99} for an SNINS structure with
perfect NS interfaces. In this case, the tunnel barrier does not affect the
Andreev reflection but produces renormalization of the diffusion coefficient,
${\cal D} \rightarrow {\cal D}R_N/R_T$.

At $eV/2\Delta \ll W^{-1/2}_\varepsilon$, when $\tau_\varepsilon \ll
\tau_d(eV)$, the normal channel may be considered as a reservoir with a
certain effective temperature (depending on the details of inelastic
scattering which are beyond our model approach), and the $I$-$V$ curve
becomes structureless. At small $W_\varepsilon$, the SNS junction behaves at
two SN junctions connected in series through the equilibrium normal
reservoir.

\section{Summary} 

We have developed a consistent theory of incoherent MAR in long diffusive SNS
junctions with arbitrary transparency of the SN interfaces. We formulated a
circuit representation for the incoherent MAR, which may be considered as an
extension of Nazarov's circuit theory\cite{CircuitTheory} to a system of
voltage biased superconducting terminals connected by normal wires in the
absence of supercurrent. We constructed an equivalent MAR network which
includes a new resistive element, "Andreev resistor", which provides
electron-hole conversion at the SN interfaces. Separate Kirchhoff rules are
formulated for electron (hole) population numbers and diffusive flows. Within
this approach, the electron and hole population numbers are considered as
potential nodes connected through the tunnel and Andreev resistors with a
distributed voltage source -- the equilibrium Fermi reservoirs in the
superconducting terminals.

The theory was applied to calculation of the $I$-$V$ characteristics. The
subgap current decreases step-wise with decreasing applied voltage in
junctions with resistive interfaces, while in junctions with transparent
interfaces an appreciable SGS appears only on the differential resistance. In
all cases, $dV/dI$ exhibits sharp structures whose maximum slopes correspond
to the gap subharmonics, $eV_n=2\Delta/n$. The amplitude and the shape of SGS
oscillations strongly depend on the interface/normal-metal resistance ratio
$r=R_{SN}/R_N$ and reveal a noticeable parity effect: difference of the
amplitudes of the even and odd structures. This effect is specific for
diffusive junctions: it comes from the strongly enhanced probability of
Andreev reflection at zero energy. Inelastic scattering results in smearing
of the SGS, which disappears at small applied voltage.

Our theory of incoherent MAR is valid as soon as the applied voltage $eV$ is
much larger than the Thouless energy: in this case, one may neglect the
overlap of the proximity regions near the NS interfaces. In the opposite
case, $eV \lesssim E_{\mathrm{Th}} \ll \Delta$, the problem of the low-energy
part of the effective circuit should be considered more carefully, by taking
into account the interference between the proximity regions. Aspects of this
ac Josephson regime have been considered in Ref.\ \onlinecite{Argaman1} in
terms of adiabatic oscillations of the quasiparticle spectrum and
distribution functions, which produce nonequilibrium ac
supercurrent.\cite{Argaman2} Within our approach, this effect will introduce
an effective boundary condition for the probability currents at small energy,
which must be included into the circuit representation of MAR in energy
space.


It is useful to discuss the effect of dephasing on MAR which has not been
sufficiently investigated yet. In the present case of diffusive junction, the
dephasing is modeled by an imaginary addition $i\Gamma$ to the energy, $E
\rightarrow E + i\Gamma$. It is easy to see that this model leads to more
pronounced SGS. Indeed, as it follows from Eq.\ (\ref{BoundaryTheta}), the
effect of large dephasing rate $\Gamma$ is similar to the effect of an opaque
interface barrier (the transmissivity parameter $W$ becomes effectively
small). This is easily understood since the dephasing suppresses the
anomalous Green's function within the normal metal similar to the effect of
the interface barrier. Thus, even for highly transmissive interfaces, the
$I$-$V$ curves for large $\Gamma \gg E_{\text{Th}}$ become similar to the
ones in the case of low-transparent interfaces (see Fig.\ \ref{SimpleIVC}),
with deficit current and pronounced SGS. This conclusion is also supported by
direct numerical calculation. It is of interest that the parity effect in SGS
almost disappears due to the cutoff of the long-range proximity effect at
small energies. In such case, the incoherent MAR regime persists at
arbitrarily small voltages and our theory is valid until the quasiparticle
relaxation will affect MAR regime as described in Sec. VIII.

\acknowledgements{ 
\vspace{-2.5mm}

Support from NFR, KVA and NUTEK (Sweden), from NEDO (Japan), and from
Fundamental Research Foundation of Ukraine is gratefully acknowledged.}

\appendix

\section*{} 
\vspace{-2.5mm}

The analytical expressions for bare conductivities $G_\pm$ and proximity
corrections $m_\pm$ can be obtained in the case of low-transparent NS
interface, $W \ll 1$, by making use of a perturbative solution of Eq.
(\ref{BoundaryTheta}):
\begin{equation} \label{Perturb}
\theta_N(E) = W\sqrt{i\Delta/E} \sinh\theta_S(E),
\end{equation}
\begin{mathletters}
\begin{eqnarray} \label{PerturbG}
R_{SN} G_\pm = {E\Theta(E -\Delta)\over \sqrt{E^2-\Delta^2}}
-{W\Delta^2\Theta[\pm(E-\Delta)]\over E^2-\Delta^2} \nonumber \\ \times
\left[ \sqrt{\Delta \over 2E} -  {W\Delta\over \sqrt{\Delta^2 -
E^2}}\left(\begin{array}{ccc} 0\\ 1 \end{array} \right) \right],
\end{eqnarray}
\begin{eqnarray} \label{Perturbm}
{m_\pm\over R_N} = {\xi_\Delta W^2 \Delta^2 \over 2d|E^2
-\Delta^2|}
\left({\Delta\over 2E}\right)^{3/2}
 [\pm 2
\nonumber \\
+\Theta(E -\Delta)-\Theta(\Delta -E)],
\end{eqnarray}
\end{mathletters}
where $\Theta(x)$ is the Heaviside function and $E$ is assumed for brevity to
be positive. From Eq.\ (\ref{PerturbG}), we obtain approximations for the
tunnel and Andreev resistances: 
\begin{mathletters} \label{ApproxR}
\begin{equation}     \label{ApproxRT}
R_T(E) \approx R_{SN} \left(1 - \Delta^2/E^2\right),
\end{equation}
\begin{eqnarray}     \label{ApproxRA}
R_A(E) \approx {2R_{SN} |E^2 - \Delta^2|\over W\Delta} \sqrt{ 2E \over
\Delta} \nonumber \\ \times\left[ 1 + W\Theta(\Delta - E) \left| 2E\Delta/
(E^2 - \Delta^2) \right|^{1/2} \right].
\end{eqnarray}
\end{mathletters}

In the vicinity of the gap edges, $|E-\Delta| \lesssim W^2$, and at small
energies, $E \lesssim W^2$, where $\theta_N(E)$ in Eq.\ (\ref{Perturb})
diverges, the following approximate solutions of Eq.\ (\ref{BoundaryTheta})
have to be used instead of Eq.\ (\ref{Perturb}): 
\begin{equation} \label{Improve2}
\exp(\theta_N) = u^2(t), \; t = 2|E-\Delta|/\Delta W^2
\end{equation}
at $|E - \Delta| \ll \Delta$, where $u(t)$ is the solution of a cubic
equation $u^3-u=\sqrt{ i/t}$, and
\begin{equation} \label{Improve1}
\sinh{\theta_N\over 2} = -{i\over \sqrt{2}} \exp \left(
-\mathop{\mathrm{Arcsinh}} \sqrt{-iE/ 2 W^2\Delta} \right)
\end{equation}
at $|E| \ll \Delta$. The asymptotics of $m_\pm = \pm R_N (\xi_\Delta/d
)\mu_\pm$ and $G_\pm$ near these ``dangerous'' points, are given by
%
\begin{mathletters}
\begin{equation} \label{Improve3}
R_{SN}G_+ = \left(\sqrt{3}t^{-1/6}/ 2W\right)\Theta(E-\Delta),\nonumber \\
\end{equation}
\begin{equation}
R_{SN}G_-\! =\! (t^{-5/6}/4W)\! \left[ \sqrt{3} \Theta(E\! -\!\Delta) \!+\!
\Theta(\Delta\!-\!E) \right],
\end{equation}
\begin{equation}\label{mupm}
\mu_+\! =\! \sqrt{2 / 3t^{1/3}}\left(\sqrt{2\!+\!\sqrt{3}}\!+\!
\sqrt{2\!-\!\sqrt{3}} \right), \; \mu_- =\sqrt{2}
\end{equation}
\end{mathletters}
at $t \ll 1$. At $E \ll W^2$,
\begin{equation} \label{Improve4}
R_{SN}G_- = 1, \; \mu_- = \left(\sqrt{2}-1\right)\sqrt{\Delta/ E},
\end{equation}

For perfect SN interface, $G^{-1}_\pm = 0$, the asymptotics of $\mu_-(E)$ at
$E-\Delta \ll \Delta$ and $E \ll \Delta$ are given by Eqs.\ (\ref{mupm}),
(\ref{Improve4}) respectively, whereas $m_+(E)$ diverges at the gap edge as
$[\Delta/2(E-\Delta)]^{1/4}$. Several examples of these dependencies
calculated numerically are presented in Fig.\ \ref{BareResistances} (see
discussion in Sec.\ IV).


\begin{references}
\bibitem{footnote} Here we apply the term ``Andreev resistance'' (and further
``Andreev conductance'') to denote particular values of $R_A$ (or $G_A$).
This must be distinguished from the term ``Andreev conductance'' widely used
in literature for physical conductance of the junction at the subgap applied
voltages.
\bibitem{negative} In the terms of the circuit theory, this means that the
``voltage drop'' between the electron and hole nodes is directed
against the probability current flowing through the Andreev resistor.
\end{references}
\end{document}